\documentclass[12pt]{article}
\pdfoutput=1

\usepackage{graphicx}
\usepackage{psfrag}
\usepackage{enumerate}

\newcommand{\be}{\begin{equation}}
\newcommand{\ee}{\end{equation}}
\newcommand{\beq}{\begin{eqnarray}}
\newcommand{\eeq}{\end{eqnarray}}

\def\[{\left [}
\def\]{\right ]}
\def\({\left (}
\def\){\right )}

\def\r2{\sqrt{2}}

 \def\simleq{\; \raise0.3ex\hbox{$<$\kern-0.75em
      \raise-1.1ex\hbox{$\sim$}}\; }
   \def\simgeq{\; \raise0.3ex\hbox{$>$\kern-0.75em
      \raise-1.1ex\hbox{$\sim$}}\; }

\newcommand{\eqref}[1]{(\ref{#1})}

\newcommand{\bbibitem}[1]{\bibitem{#1}\marginpar{#1}}

\newcommand{\figref}[1]{Fig. \ref{#1}}

\def\Label#1{\label{#1}%
  \smash{\hbox to0pt{\raise1ex\hbox{\tiny[#1]}\hss}}}
\def\noLabels{\let\Label=\label}
\def\nobbibitem{\let\bbibitem=\bibitem}

\begin{document}

\noLabels
\nobbibitem

\DeclareGraphicsExtensions{.pdf,.png,.gif,.jpg,.eps}

\begin{titlepage}

\begin{center}
{\Large \bf Cosmic Bubble Collisions}
\vspace{6mm}

{Matthew Kleban}

\vspace{6mm}
{\it Center for Cosmology and Particle Physics\\
New York University \\
New York, NY 10003, USA}

\end{center}

\begin{abstract}
\noindent

I briefly review the physics of  cosmic bubble collisions in false-vacuum eternal inflation.  My purpose is to provide an introduction to the subject for readers unfamiliar with it, focussing on recent work related to the prospects for observing the effects of bubble collisions in cosmology.  I will attempt to explain the essential physical points as simply and concisely as possible, leaving most technical details to the references.  I make no attempt to be comprehensive or complete.  I also present a new solution to Einstein's equations that represents a bubble universe after a collision, containing vacuum energy and ingoing null radiation with an arbitrary density profile. 

\end{abstract}
\end{titlepage}







\tableofcontents
	
\section{Introduction}

This work reviews the physics of cosmic bubble formation and collisions, with a focus on recent work.
It will be as self-contained as possible while avoiding technical details, with references to the literature where further details can be found.  I will use natural units ($8 \pi G = c = \hbar = 1$) throughout.  Another review of cosmic bubble collisions is \cite{ajreview}.

\subsection{Overview}

First-order phase transitions are ubiquitous in physics.  During a first-order transition, a meta-stable phase (the ``false'' vacuum) decays to a lower energy phase (the ``true'' vacuum, which may itself be either metastable or truly stable) either by quantum tunneling or because the decay is stimulated by some external influence.  The transition begins in a finite region---a bubble.  If the bubble is sufficiently large when it forms, it  expands into the false vacuum and collides with other bubbles.  Therefore in a static spacetime, the transition will eventually percolate and the false vacuum will disappear entirely.

\begin{figure}
\hspace{0 in}\includegraphics[width=1\textwidth]{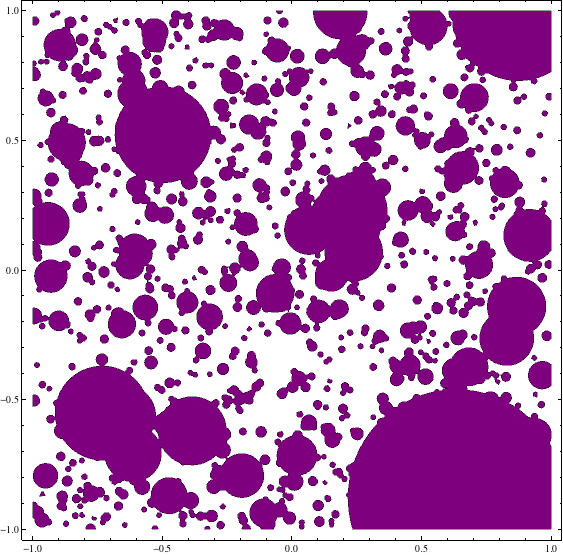}
\caption{\label{fractal} A numerical simulation showing the spatial distribution of bubbles at a late time in an eternally inflating false vacuum.  Bubbles that appeared earlier expanded for longer and are larger, but the physical volume of the false vacuum is larger at later times, so there are more smaller bubbles.  Eventually, each bubble collides with an infinite number of others.  The late-time distribution is a scale-invariant fractal.}
\end{figure}

The situation is different when gravity is included.  The energy density in a meta-stable phase does not change with the expansion of space; it is a cosmological ``constant''.  Therefore, if the false vacuum has positive energy it will undergo cosmic inflation---the volume of space containing it will expand exponentially with time.  If the timescale for the exponential (the false-vacuum Hubble rate) is faster than the rate of bubble formation, the false vacuum will continually reproduce itself and the transition will never percolate.  Bubbles of true vacuum will expand (but slower than the false vacuum) and occasionally collide with each other (Fig. \ref{fractal}).  Contained inside each bubble is an expanding Friedmann-Robertson-Walker (FRW) cosmology that is  homogeneous and isotropic apart from random perturbations and the aftereffects of collisions.

In  models consistent with current observations, our observable universe is  inside such a bubble, embedded in and expanding into an eternally inflating parent false vacuum.  Observational constraints require that our bubble underwent its own period of slow-roll inflation after its formation.  Collisions with other bubbles that nucleate in the (otherwise eternal) parent false vacuum nearby occur with a non-zero probability per unit time, and are therefore guaranteed to happen eventually.  Their effects have already or will perturb the universe around us, creating potentially observable signals.  In terms of the FRW coordinates describing the cosmology inside our bubble these collisions occur {\it before} slow-roll inflation---in fact, they occur before the (apparent) big bang of our FRW universe.  They can be regarded as creating a special and predictable set of initial conditions at FRW time $t=0$.

\subsection{Effective field theory coupled to gravity}

Any model of spacetime fields coupled to gravity can give rise to bubble collisions if there exist at least two meta-stable phases of the field theory.   After forming, bubbles expand and  collide with each other.  The goal of this review is to describe the physics of the formation, expansion, and collision of these bubbles.  I will focus exclusively on models in which at least one of the phases (the false vacuum) has a positive vacuum energy.   

A region of spacetime filled with positive vacuum energy has a metric 
\be \label{frwmet}
ds^2 = -dt^2 + a(t)^2 d\vec x^2
\ee
and obeys Einstein's equations 
\be \label{FRW}
(\dot a/a)^2=H_f^2 = {V_f / 3},
\ee
where $V_f$ is the energy density of the false vacuum and $H_f$ is the associated Hubble constant.  The solution to \eqref{FRW} is de Sitter space, $a(t)=e^{H_f t}$.  Regions of space that are dominated by vacuum energy but contaminated by other forms of matter or energy will exponentially rapidly inflate away the contaminants and approach the metric \eqref{frwmet}.  Regions not dominated by vacuum energy will either expand more slowly or collapse into black holes, which for many purposes is taken as justification for ignoring them after a few false vacuum Hubble times (where the Hubble time is $t_{H} \equiv 1/H$).

\subsection{Decay}

In metastable de Sitter space there is a dimensionless rate of bubble formation $\gamma$.   When $\gamma$ is small it can be defined as the expected number of bubble nucleations per unit Hubble time per unit Hubble volume:  that is, the dimensionful decay rate is $\Gamma=H_f^4 \gamma$.  Generally $\gamma$ is the exponential of $-S$, where $S$ is the action for an instanton.  Hence, when $S \gg 1$, $\gamma$ is very small and the rate of bubble nucleations is  slow.  When $\gamma \simgeq 1$, the semi-classical methods  reviewed here are not adequate to describe the physics.

When $\gamma$ is small and the meta-stable phase has positive vacuum energy, the exponential expansion means that the transition will never percolate---there will always be some regions in which the unstable phase remains.  The intuitive reason is simple:  in one Hubble time, the de Sitter region increases its volume by a factor of $e^{3}$.  In that same Hubble time, one expects $\gamma$ bubbles of another vacuum, each of volume less than the Hubble volume, to appear.  Therefore if $\gamma \ll 1$ a typical region will double in size many times before producing a bubble.  The same applies to each ``child" region; and therefore the meta-stable phase never ceases to exist.  This is known as ``false vacuum eternal inflation" (to distinguish it from slow-roll eternal inflation, which can take place in an models with sufficiently flat, positive potential energy functionals).

In the end, the picture is an exponentially rapidly expanding spacetime in which bubbles of more slowly expanding phases occasionally appear, and occasionally collide.  It is important to note that this is a generic prediction of any model with multiple, positive energy minima---while it seems to be a prediction of the string theory landscape, it is certainly not unique to it.  Nevertheless, an observation that confirmed this model would be a confirmation of a prediction of string theory, and an observation that ruled it out would be at least potentially a falsification.

\subsection{Motivation} \label{mot}

In the last few years there has been a surge of interest in this problem.  The reason is that string theory predicts the existence of many
meta-stable minima, the so-called ``string landscape" \cite{discretum, lennylandscape}.  In string theory, the geometry and topology of spacetime is dynamical.  String theories exist in 9 spatial dimensions.  Since we observe only three spatial dimensions, in string solutions that might describe our world six of the spatial dimensions are compactified; that is, they form a geometry with finite volume, while the other three spatial dimensions and time can form Minkowski or de Sitter space.  

Six dimensional manifolds have many parameters that describe their shape and size.  These parameters are dynamical fields in string theory, and their meta-stable solutions correspond to distinct possibilities for the shape and size of the manifold.  Because they are meta-stable, small fluctuations around these geometries behave like massive particles from the point of view of the 4 large spacetime dimensions.  Therefore the low-energy physics as measured by a 4D observer is an effective field theory coupled to gravity, with the field content partially determined by which configuration the compact manifold is in.

The compact manifold can make transitions from one meta-stable configuration to another.   
Among the parameters that can vary from configuration to configuration is the value of the vacuum energy.  In non-supersymmetric solutions it will not be zero; instead, its typical value is set by the string energy scale---with both positive and negative values possible \cite{discretum, kklt, lennylandscape}.   The exact value in any given phase will depend on the details, and if there are enough different phases, it is plausible that a small but non-empty subset have vacuum energies that are consistent with the tiny value we observe \cite{riess, perlmutter, Komatsu:2010fb}.  Hence the theory is at least consistent with observation, and the extraordinarily small but non-zero value was even predicted a decade in advance in \cite{w} under a set of assumptions that coincide with those just outlined.

Because understanding the value of the cosmological constant is considered one of the most important problems in theoretical physics, and because string theory is our best candidate for a theory of quantum gravity, it is worth taking this scenario seriously.  One should look for  observations beyond the cosmological constant which could test its validity; hence this review.

\section{Earlier work}

Alan Guth's original model for inflation---now known as ``old inflation''---was an inflating false vacuum punctuated by bubbles \cite{guth}.  Inflation was the false vacuum, and  the end of inflation was the  nucleation of the bubble we inhabit.  This model didn't succeed because of difficulties with re-heating the universe:  when a bubble first appears it is both strongly negatively curved and empty.  Reheating could potentially result from collisions with other bubbles, but such collisions are either rare---in which case reheating is anisotropic---or common---in which case the phase transition from the false vacuum completes after order one Hubble time, and there is little or no inflation.  Early studies of this include \cite{gw, hms}.  

\subsection{Open inflaton}

Old inflation was replaced by other models in which the inflaton slowly rolls, and less attention was paid to models involving bubbles until the mid-90s when the available data indicated that the universe was underdense and negatively curved.  At that time there was a burst of interest in ``open inflation'' models, models in which our universe is inside a bubble that nucleated from a parent false vacuum (and hence is negatively curved), but where the bubble nucleation was followed by a period of standard slow-roll inflation.  That period of slow-roll produces density perturbations and reduces the curvature, thereby allowing $\Omega_{\rm total}\sim.3$ as the data seemed to indicate.  There were a series of papers by several groups of authors that worked out the power spectrum of perturbations generated during inflation in such bubbles
\cite{Bucher:1994gb, Linde:1995rv, White:1996zi, Yamamoto:1996qq, Linde:1999wv}).

While not all authors agreed on the quantitative effects, all agree that the power spectrum of inflationary perturbations is significantly modified on scales of order the radius of curvature and larger.  This can be understood qualitatively in a simple way.  As I will review below, when a bubble universe first appears it is dominated by negative curvature.  As a result it expands with a scale factor $a(t)\sim t$.  After a time of order $H_i^{-1}$, where $H_{i}$ is the Hubble constant during inflation, the curvature redshifts to the point it is below the inflationary vacuum energy and inflation begins.  Clearly, perturbations produced before or during this transitional era will not have a scale-invariant spectrum identical to those produced after it.

Mainly because of theoretical developments in string theory (see Sec. \ref{mot}), in the last few years there has  been a resurgence of interest in these models, focussing both on the physics of individual bubble universes and on collisions between bubbles \cite{Garriga:1998px, fkms, ggv, fhs, Aguirre:2007an, wwc,Aguirre:2007wm, wwc2, bubmeas, Aguirre:2008wy, ktflow,Easther:2009ft, Giblin:2010bd, Czech:2010rg, Salem:2010mi}.  I will review much of that very recent work below.  A  result from some years back is \cite{Garriga:1998px, fkms}, which demonstrate in several models (using  arguments parallel to \cite{w}) that slow-roll inflation after the bubble nucleation is necessary in order to avoid producing a universe devoid of structure.  With fixed $\delta \rho/\rho$,  approximately as many efolds of inflation $N$ are needed as are necessary to solve the flatness problem (approximately $N\sim 60$ given typical values for the reheating temperature, etc.).

\subsection{Metrics and solutions}

The single most important tool in understanding the physics of cosmic bubbles is their symmetry.  The standard treatment of bubble formation in field theory coupled to gravity was developed by \cite{cdl}.  In that approach, both the rate of nucleation and the initial conditions just after the bubble forms are set by a solution to the equations of motion of the Euclidean version of the theory.   Such solutions are known generally as ``instantons". The justification for their application to this problem goes far beyond the scope of this review;  see \cite{cdl} and references therein.

Another mechanism by which false vacua can decay is charged membrane nucleation in a background 4-form flux \cite{Brown:1988kg}.  These instantons (and the expanding bubbles they correspond to) possess the same symmetries as those of \cite{cdl}, and I expect most of the results reviewed here to hold for them as well.

In order to understand the symmetries of the bubble spacetime, one can start with the instanton solution of \cite{cdl}.  In a model with a single scalar field $\phi$ coupled to Euclidean Einstein gravity, the relevant solution is
\be \label{eucmet}
ds^2 = d \psi^2 +  a(\psi)^2 d \Omega_3^2, ~~~~~~~~~~~~ \phi = \phi(\psi),
\ee
where $d \Omega_3^2 = d \theta^2 + \sin^2 \theta d\Omega_2^2$ is the round metric on a 3-dimensional sphere.  The functions $ a(\psi)$ and $\phi(\psi)$ are determined by solving Einstein's equations and the equation of motion for the scalar field $\phi$, and the solutions depend on the potential energy for the scalar $V(\phi)$.  If $V$ has at least two minima, solutions  $\phi(\psi)$ that oscillate once around the maximum\footnote{In Euclidean signature, the equations of motion have the sign of the potential reversed.}  of $V(\phi)$ that separates the two minima generally exist (but see \cite{hm} and \cite{pk}). 
 
Solutions of this form have a large degree of rotational symmetry:  they are invariant under the group of rotations in 4D Euclidean space (in which one can embed the 3-sphere), namely $SO(4)$.  This is a group with $4\times 3/2=6$ symmetry generators (each corresponding to a rotation in one of the six planes of 4D space).  The action $S(a, \phi)$ for a solution is determined by an integral involving $a(\psi)$ and $\phi(\psi)$, and is related to the false vacuum decay rate by $\gamma \sim e^{-S}$.

Under certain conditions on the potential $V(\phi)$, the solutions $a, \phi$ are ``thin wall", meaning that they vary sharply with $\psi$.  In particular, in the thin wall limit $\phi(\psi)  \sim \phi_0 + \mu \Theta(\psi-\psi_{0})$, where $\Theta(\psi)$ is a step function and $\phi_0, \mu$, and $\psi_0$ are constants.  The thin wall limit is convenient; it makes it possible to calculate many quantities in closed form.  However it is {\it not} required for the validity of most of the physics discussed in this review, nor do the symmetries of the solutions discussed here depend on it.

To determine the spacetime after the bubble nucleation occurs, one can simply analytically continue \eqref{eucmet} back to Lorentzian signature.  The result is a spacetime that contains a bubble expanding in a bath of false vacuum.  There is more than one choice of analytic continuation; different choices give metrics that cover some part of the full Lorentzian spacetime.  These metrics can be patched together  to determine the global structure of the spacetime.  Of most relevance for the cosmology seen by observers inside the bubble is the continuation that produces a metric that covers (most of) the interior of the bubble.  Details can be found in {\it e.g.} \cite{fkms}; the result is
\be \label{frw}
 ds^2 = -dt^2 + a(t)^2 d H_3^2=-d t^2 + a(t)^2 \( d\rho^2 + \sinh^2 \rho d\Omega_2^2 \) ~{\rm and}~ \phi = \phi(t),
\ee
where $H_3$ is 3D hyperbolic space (the homogeneous and isotropic 3D space with constant negative curvature).  As can be seen at a glance, this is an FRW metric describing a homogeneous and isotropic cosmology scale factor $a(t)$ and scalar field ``matter" $\phi(t)$.  The evolution of $a$ will be determined by $V(\phi(t))$ in the usual way, as well as by any other sources of stress-energy.  

After the continuation, the $SO(4)$ invariance of the Euclidean solution has been replaced by the $SO(3,1)$ invariance associated with $H_3$.  In other words, the symmetries of the instanton are what give rise to the homogeneity and isotropy of the constant-time surfaces of this FRW cosmology.  Observers living inside the bubble live in a negatively curved universe that satisfies the cosmological principle.

One of the most intriguing aspects of this result is that the ``big bang" $t=0$, where the scale factor $a(0)=0$ vanishes, is {\it not} a curvature singularity.  It is merely a {\it coordinate} singularity---the spacetime just outside the surface $t=0$ (which is null) is regular and described by another analytic continuation of the metric.  Of course there is a quantum nucleation event (a finite distance outside the surface $t=0$) that may make the description there in terms of a classical spacetime plus field configuration problematic, but so long as the energy densities involved are sub-Planckian, even the region outside the ``big bang'' surface $t=0$ is no more singular than the nucleation of a bubble in a first-order phase transition in field theory.

\subsection{Bubblology}

The cosmology inside the bubble in the region described by \eqref{frw} is fairly simple to describe.  The scale factor obeys the Friedmann equation
\be
H^2 = (\dot a/a)^2 = \rho/3 + 1/a^2,
\ee
where $\rho \sim V(\phi) + (\dot \phi)^2/2+...$ is the energy density in the scalar plus that of any other matter or radiation.  At $t=0$, the initial conditions are such that $\dot \phi = 0$.  As a result, the dominant term on the right hand side is the curvature term $1/a^2$, and one  obtains $a(t) = t + {\cal O}(t^3)$ \cite{fkms}.

\begin{figure}
\hspace{0 in}\includegraphics[width=1\textwidth]{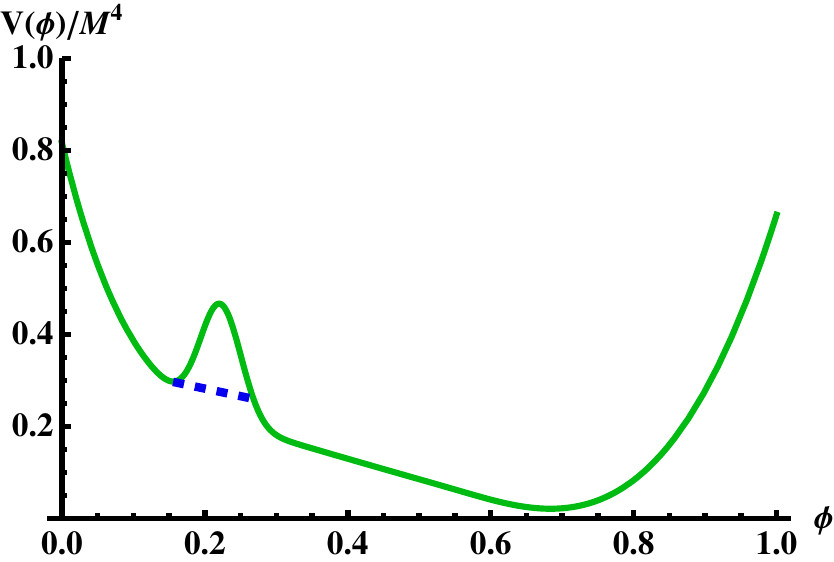}
\caption{\label{potential} A scalar field potential for an open inflation model, where a single field $\phi$ tunnels through a barrier and then drives slow-roll inflation inside the resulting bubble.}
\end{figure}

To produce a cosmology consistent with observations, there must be a period of slow-roll inflation that begins at some time $t>0$ \cite{fkms}.  The most economical way to accomplish this is to assume that the field that tunneled to produce the bubble in the first place is also the inflaton, and that its potential $V(\phi)$ has a slow-roll ``plateau" that the field evolves into after the tunneling (see Fig. \ref{potential}).

In such a model, inflation begins at a time $t_i$ such that $1/a(t_i)^2 \sim 1/t_i^2 \sim V(\phi(t_i)) \sim H_i^2$.  Therefore, inflation begins at $t=t_i \sim 1/H_i$.  From that time on the scale factor will begin to grow exponentially, rapidly diluting the curvature and (given enough efolds) solving the curvature problem.

\subsection{Curvature and fine-tuning}

During inflation, the radius of spatial curvature of the universe grows exponentially.  In terms of the total energy density $\Omega$, the curvature contribution to the Friedmann equation is defined by $\Omega_{k} \equiv 1-\Omega=-k /(H a)$, where $k=-1$ for negative curvature.  As described above, $\Omega_{k} \simleq 1$ at the start of slow-roll inflation.  Since during inflation $H$ is approximately constant and $a$ grows exponentially, by the end of inflation after $N$ efolds $\Omega_{k} \sim e^{-2 N}$.

After inflation, $\Omega_{k}$ grows by a very large factor $e^{2 N_{*}}$, so that its value today is $\Omega_{k}\sim e^{2(N_{*}-N)}$ ($N_{*}\sim 63$ depends logarithmically on the reheating temperature and various other factors; see {\it e.g.} \cite{ll}).  

Therefore, in open inflation models the value of $\Omega$ today is exponentially sensitive to the length of inflation.  As with nearly all relics of the state of the universe prior to the start of inflation, inflation ``inflates away'' curvature with exponential efficiency.  

This raises the question of how fine-tuned an open inflation model with observable curvature would be.  The answer depends on the expected number of efolds of slow-roll inflation.  As mentioned  above, \cite{fkms} demonstrated that too little inflation ($N<N_{*}$) prevents structure formation, at least given a fixed ($N$-independent) value of $\delta \rho/\rho$.  The argument parallels that of Weinberg  \cite{w}.  Negative curvature can be thought of as a velocity for the expansion of the universe that exceeds ``escape velocity''.  As such, overdensities collapse only if the overdensity exceeds a certain bound---in effect, in order to collapse an overdense region must be dense enough that it is locally a closed universe.  Therefore, for a given amplitude of $\delta \rho/\rho$, insufficient $N$ makes collapsed structures ({\it i.e.} stars and galaxies) exponentially rare.

Following Weinberg's logic \cite{w}, one may therefore expect that $N$ is not much greater than $N_*$.  How much greater we expect it to be is a strongly model-dependent question: in \cite{fkms} a toy model gave a measure of for the probability of $N$ efolds $P(N) \sim N^{-4}$.

\section{Collisions}

In any given  Lorentz frame, every bubble must eventually  undergo a first collision.  In this section I will discuss the effects on cosmology of a single collision---strictly speaking, the analysis does not apply once there are regions affected by multiple collisions.  However if the effects are sufficiently weak in those regions, one can use linear cosmological perturbation theory, in which case the effects of collisions simply superpose. 

\begin{figure}
\hspace{0 in}\includegraphics[width=1.8 \textwidth]{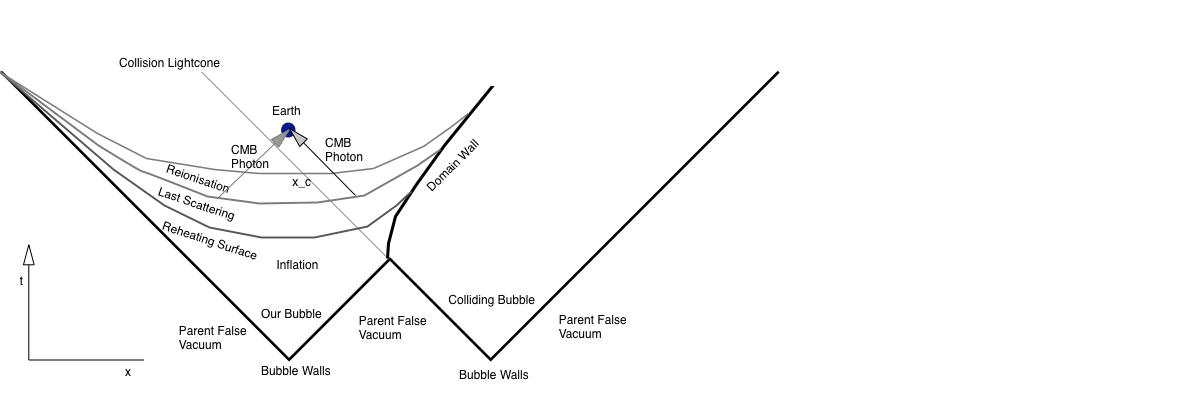}
\caption{\label{spacetimef}A spacetime diagram showing the causal structure of a cosmic bubble collision.  Coordinates are chosen so that light propagating in the plane of the diagram moves along $45^{{\circ}}$ lines.}
\end{figure}

When two bubbles collide, the spacetime region to the future of the collision is affected (see \figref{spacetimef} for a spacetime diagram of a bubble collision).  Precisely what occurs inside that region depends on the model.  However for applications to observational cosmology, one is generally interested in regions in which the effects are a small perturbation on the spacetime prior to the collision.  As I will discuss below, the assumption that the perturbation is small plus the symmetries of the collision suffice to extract the generic leading-order effects on cosmology.

The most important features of a collision of this type are related to its symmetries.  As reviewed above, a single bubble has an $SO(3,1)$ isometry, a group generated by 6 symmetry generators.  It turns out that a bubble collision breaks the $SO(3,1)$ to $SO(2,1)$, a group with 3 generators \cite{hms}.  This is enough symmetry to solve Einstein's equations in a situation in which the stress tensor is locally dominated by vacuum energy---the situation is very similar to that of a black hole (with three rotation isometries) in a vacuum dominated spacetime (de Sitter, Minkowski, or anti-de Sitter).  As in those examples, there is a one parameter family of solutions.  Details can be found in {\it e.g.} \cite{hms, fhs, wwc, Aguirre:2007wm}.

We will see that as a result of these symmetries, the effects of the collision break isotropy and homogeneity but are azimuthally symmetric (and non-chiral)  around a special direction---the axis pointing toward the center of the colliding bubble.

One important feature of these symmetries is that there exists a time slicing of the collision spacetime in which the entire collision occurs at one instant everywhere along a 2 dimensional hyperboloid.  
At later times in these coordinates, the effects of the collision spread along a light ``cone'', which is really a region of space bounded by two hyperboloids.  To visualize this, it may help the reader to visualize the intersection of two lightcones cones or ``hourglass-type'' timelike hyperboloids in 2+1 dimensions.  The intersection is a (spacelike) hyperbola, which can be chosen to correspond to an instant of time and a point in one transverse coordinate ({\it e.g.} $t$ and $x$ in \eqref{bc}).

\subsection{Thin-wall collision metric}

In the approximation that the spacetime everywhere away from thin surfaces is dominated by vacuum energy, one can write down exact solutions.  These metrics are of the form
\be \label{bc}
ds^2 = -dt^2/g(t) + g(t) dx^2 + t^2 dH_2^2,
\ee
where $dH_2^2=d \rho^2 + \sinh^2 \rho d\phi^2$ is the metric on a 2D hyperboloid, and 
\be
g(t) = 1 + H^2 t^2 - m/t.
\ee
Here $H$ is the Hubble constant related to the local vacuum energy density $\rho_V$ by $H^2 = \rho_V/3$, and $m$ is a constant \cite{hm, fhs, wwc}.  With $m=0$ this is simply de Sitter space.  But with $m$ non-zero, it represents a solution that is essentially an analytically continued Schwarzschild-de Sitter black hole (see \cite{wwc} for a discussion of the causal structure of these solutions).

Across a wall that separates two vacua (either the wall that separates two collided bubbles, or the walls of the bubbles themselves in the parent false vacuum), one expects the vacuum energy density $\rho_V$ to jump, along with many other quantities.  In the limit the wall is thin, regions described by \eqref{bc} can be ``glued" together using Israel matching conditions \cite{Israel:1966rt, israelnull, hms, fhs}.  This allows one to determine the trajectory of the domain walls, and therefore the conditions under which the domain walls that form in a collision accelerate towards or away from inertial observers inside the bubbles.  When the wall accelerates, it rapidly becomes relativistic.  Anything it collides with will almost certainly not survive, as the ultra-relativistic wall typically has a very high energy density even in its rest frame.  In such cases there are no cosmological observables, since the wall arrives almost immediately after its first light signal.  

The results of this analysis, which hold at least when the spacetime on either side of the domain wall is dominated by vacuum energy, are as follows:   the domain wall always accelerates away from an inertial observer in the bubble with lower vacuum energy, and---depending on the tension in the wall and the difference in the two vacuum energies---it can either accelerate towards, away from, or not accelerate with respect to an observer in the bubble with larger vacuum energy \cite{wwc, Aguirre:2007wm}.   One interesting implication of this result is that inertial observers in bubbles with small positive vacuum energy are shielded from the walls separating them from regions with higher vacuum energy, and might be shielded from walls with negative vacuum energy beyond them if the potential satisfies certain conditions.

Generalizing the solution \eqref{bc} to the case of non-vacuum energy dominated spacetimes is difficult---roughly as difficult as embedding a black hole into a non-trivial FRW cosmology.  However the basic symmetry structure will not be affected:  the metric must still posses $SO(2,1)$ symmetry and therefore can be written in the form $ds^2 = ds_2^2 + f(t,x) dH_2^2$ (just as a black hole embedded in an expanding FRW could be written in a form $ds^2 = ds_2^2 + f(t,r) d\Omega_2^2$.  In particular, the region affected by the collision---which takes place at some definite values of $t$ and $x$---is still bounded by the future light ``cone" of the hyperboloid $H_2$ at $t,x$.

\subsection{A new solution}

The metric \eqref{bc} is a specific analytic continuation of an anti-de Sitter Schwarzschild black hole metric.  There exist generalizations known as de Sitter-Vaidya or anti-de Sitter-Vaidya metrics that represent the formation of a black hole in de Sitter or anti-de Sitter spacetime from the collapse of a shell of null radiation \cite{Vaidya:1951zz, Wang:1998qx}.  The shell does not have to be thin---in fact these metrics contain a free function that describes the profile of the ingoing radiation.  The only criterion is that the radiation be purely radially ingoing (or purely outgoing).

Performing the appropriate analytic continuation, one  obtains the following metric:\footnote{I thank T. Levi and S. Chang for discussions on this metric.}
\be \label{vaid}
ds^2 =  (1- m(w)/t + H^2 t^2) dw^2 - 2dw dt + t^2 dH_2^2,
\ee
If $m(w)=m$ is constant,  \eqref{vaid} is equivalent to \eqref{bc} by a coordinate transformation.  In general, it is easy to check that it represents a spacetime with $SO(2,1)$ isometries that contains both vacuum energy (with energy density determined by $H$ as above) and radiation (with energy flux proportional to $m'(w)$)---but with all the radiation moving coherently in the same direction.  As such, it may be a fairly good approximation to the spacetime inside a bubble after a collision, at least during the curvature domination and slow-roll inflation phases.  To my knowledge this is the first time this solution has appeared in print.

\section{Cosmological effects of collisions}

To determine the effects of the collision on cosmological observables, I will make the following assumptions:

\begin{enumerate}
\setcounter{enumi}{-2}

\item We are inside a bubble that has been or will be struck by at least one other bubble.

\item The collision preserves $SO(2,1)$; that is, one can choose coordinates like \eqref{bc} or \eqref{vaid} in which it occurs at one instant everywhere along a hyperboloid.

\item The effects of the collision are confined entirely to the interior of the future ``lightcone'' of the collision event; that is, inside a region $x>x_c(t)$, where $x_{c}(t)$ is a null geodesic in the $(x,t)$ plane transverse to the collision hyperboloid.

\item There was sufficient inflation to make the spatial curvature of the universe and of the hyperbolic collision ``lightcone'' negligible.

\item The effects on the observable part of our universe are small enough to be treated using linear perturbation theory.

\item The collision affects the inflaton by creating an ${\cal O}(1)$ perturbation in it near the beginning of inflation.

\item The inflaton perturbation is generic, apart from constraints imposed by the symmetries and assumption 1.

\end{enumerate}

Using these assumptions, the effects on the cosmic microwave background (CMB) temperature were  first worked out in \cite{wwc2}, and on CMB polarization in \cite{Czech:2010rg}.

\subsection{Inflaton perturbation}

Given these assumptions, one can simply expand the inflaton perturbation in a power series at a time near the beginning of inflation, at which time th the metric \eqref{bc} will be a good description of the spacetime.  Dropping the $H_{2}$ factor (on which the perturbation is constant) for clarity:
\be
ds^{2} = -dt^{2}/g(t) + g(t)dx^{2} \approx -dt^{2}/(H_{i}t)^{2}+(H_{i}t)^{2}dx^{2}=(H_{i} \tau)^{-2}\(-d\tau^{2}+dx^{2} \).
\ee
The approximation is valid after the beginning of inflation $H_{i} t > 1$, and the coordinate $\tau = -1/(H^{2}t)$ is the conformal time.  The reader should recall that $\tau$ increases towards zero during inflation, and is exponentially small by its end.  

The constraint from assumption 1. means that the collision perturbation is non-zero only inside the affected region \cite{wwc2, gk}: 
\begin{eqnarray} \label{infexp}
\delta \phi(x, \tau_i) = M \( \sum_{n=0}^\infty a_n H_i^n(x+\tau_{i})^n \) \Theta(x+\tau_{i}),\\ ~~~~~ \dot {\delta \phi}(x, \tau_i) = M  \( \sum_{n=0}^\infty b_n H_{i}^{n+1}(x+\tau_{i})^n \) \Theta(x+\tau_{i})
\end{eqnarray}
Here $\delta \phi$ is the perturbation in the inflaton field, $a_n$ and $b_{n}$ are dimensionless coefficients, $\tau_i$ is a time near the start of inflation where the slow-roll approximation is valid, $M$ is a mass scale associated with the range over which the field varies, and the coordinates are chosen for convenience so that the edge of the region affected by the collision is located at $x=-\tau_i$ at that time.  Note that this expansion is completely general---it does not assume that the field that tunneled is the inflaton, and depends only on the assumptions outlined just above.  It does however ignore the effects of the collision on fields other than the inflaton; at least in single-field models one expects that primordial perturbations of the inflaton field will have the largest imprint on cosmology today.

Without a model for the inflaton and the field(s) that participated in the tunneling, we cannot compute the coefficients $a_n$ and $b_n$.  Generically one expects that the perturbation will be at least ${\cal O}(1)$ on scales corresponding to the radius of curvature of the universe.   That is, one expects the coefficients $a_n \sim 1$.  Moreover, given enough inflation to solve the flatness problem (and produce a universe consistent with current constraints on curvature), the observable universe today corresponds to a region of size  $|x| H_{i} \sim \sqrt{\Omega_{k}(t_{0})} \ll 1$ at the beginning of inflation.  Therefore, we are interested in the perturbation \eqref{infexp} in the limit $H x \ll 1$, and so the first term will dominate the observable signatures (unless for some reason $a_{0} \ll a_{n}$ for some $n>0$).  Notice that  $a_0 \neq 0$  is a spatial discontinuity in $\delta \phi$.

To determine the effects of this initial inflaton perturbation on the CMB and other cosmological observables, one can employ the standard machinery of inflationary perturbation theory.  The first step is to determine the evolution of the perturbation during inflation.  During slow-roll inflation, at lowest order in the slow-roll parameters inflaton perturbations satisfy the equation of a free, massless scalar in de Sitter space (see {\it e.g.} \cite{mfb}).  One can solve that equation in full generality: the solution is  \cite{wwc2}
\be \label{infsol}
\delta \phi = g(\tau + x) - \tau g'(\tau +x) + f(\tau-x) - \tau f'(\tau-x) 
\ee
where $g, f$ are arbitrary functions of one variable, and
$g',f'$ their derivatives.

This solution may be familiar from the more standard treatment of inflationary fluctuations in momentum space: the terms proportional to $\tau$ are the usual decaying modes in the expression $\delta \phi_k \sim k^{-3/2}\left(1 \pm ik \tau \right)$.  Evidently the $\tau$-dependent terms arise due to the effects of inflation, since this solution is otherwise identical to that of a massless scalar field in 1+1-dimensional Minkowski space.

Given the initial conditions in \eqref{infexp}, one can use \eqref{infsol} to determine  the solution for all future times during slow-roll inflation.  The result can be written in closed form \cite{gk}, but is not particularly illuminating.  Rather than reproduce it here, it is easier to instead expand the functions $f$ and $g$ of \eqref{infsol} (which contain the same information as and are determined by $\delta \phi$ and $\dot {\delta \phi}$).  Since $g$ is purely left-moving and $f$ is purely right-moving, for $\tau>\tau_i$ the perturbation near the edge of the lightcone $H_{i}(x - x_c(\tau)) \ll 1$ is determined only by $g$.  Therefore, the effects on the CMB are determined by the coefficients in the expansion of $g$:
\be
 g(x+\tau) = \( \sum_{n=1}^\infty c_n (x + \tau)^n \) \Theta(x+\tau).
\ee
Note that the sum begins with the linear term $n=1$.  This is because an $n=0$ term in this expansion would lead to a $\delta$-function singularity in $\delta \phi$, which could not be expanded as in \eqref{infexp} and in any case would invalidate perturbation theory.  Put another way, a non-zero $c_1$  corresponds to a perturbation $\delta \phi$ with a non-zero $a_0$ in \eqref{infexp}.

By the end of inflation, $\tau=\tau_e$ is exponentially small and one can drop the terms proportional to $\tau$ in \eqref{infsol}, leaving simply
\be
\delta \phi(x, \tau_e) \approx g(x) =  \( \sum_{n=1}^\infty c_n (x)^n \) \Theta(x).
\ee
Note: {\it  the leading term is now a discontinuity in the first derivative}, not a discontinuity in the perturbation itself.   Inflation smooths  sub-horizon primordial gradients by effectively integrating them once.

\subsection{Post-inflationary cosmology and Sachs-Wolfe}

To precisely determine the effect of this perturbation on cosmological observables at later times requires numerically integrating the coupled equations that control the perturbations in the various components of the energy density in the early universe.  There is a standard set of software tools available for this purpose \cite{cmbfast1, camb}.  The results of such a numerical analysis for a bubble collision can be found in \cite{kls}, and analytically (and approximately) in \cite{wwc2, Czech:2010rg}.

One can understand the qualitative results with a simple analytic approximation.  A primary component of the collision signal is its effect on the CMB temperature.  The most important contribution to CMB temperature fluctuations comes from the Sachs-Wolfe effect 
\cite{sw, wh}.  
To first approximation all CMB photons last scattered at redshift $z_{dc}\sim 1100$ when electrons recombined with protons and photons decoupled from the plasma, and then propagated freely (without scattering again) until now.  Their temperature is determined by the intrinsic temperature $T(t_{dc},x,y,z)$ at the point in the last scattering volume at the time $t_{dc}$ corresponding to $z=z_{dc}$.  Given that a photon free-streamed, it originated at some point satisfying $x^2+y^2+z^2=D_{dc}^2$ on a sphere of radius $D_{dc}$, the radius of the current earth's past lightcone at $t=t_{dc}$.  The radius $D_{dc} \approx 14$Gpc can be determined from the expansion history $a(t)$ by integrating back along a null geodesic in the FRW metric.  

Because a locally higher temperature at last scattering corresponds to a locally higher density, there is a corresponding negative gravitational potential energy fluctation at that location.  In the Sachs-Wolfe approximation the current temperature of a photon originating from such a point is a combination of the intrinsic temperature of the plasma it last scattered from and the local gravitational potential at that point, as well as an overall redshift $(1+z_{dc})^{-1}$ (and  some additional contributions from time variations of the potential along the way, the ``integrated Sachs-Wolfe effect'').  A simple calculation shows that the negative potential energy contribution is larger than the increased intrinsic temperature perturbation: $\Phi(t_dc)\sim -(3/2)\delta T/T(t_{dc}$ \cite{wh}.  Therefore, locally hot spots at last scattering make cold spots on the CMB.

\subsection{CMB temperature}

\begin{figure}
\hspace{0 in}\includegraphics[width=1\textwidth]{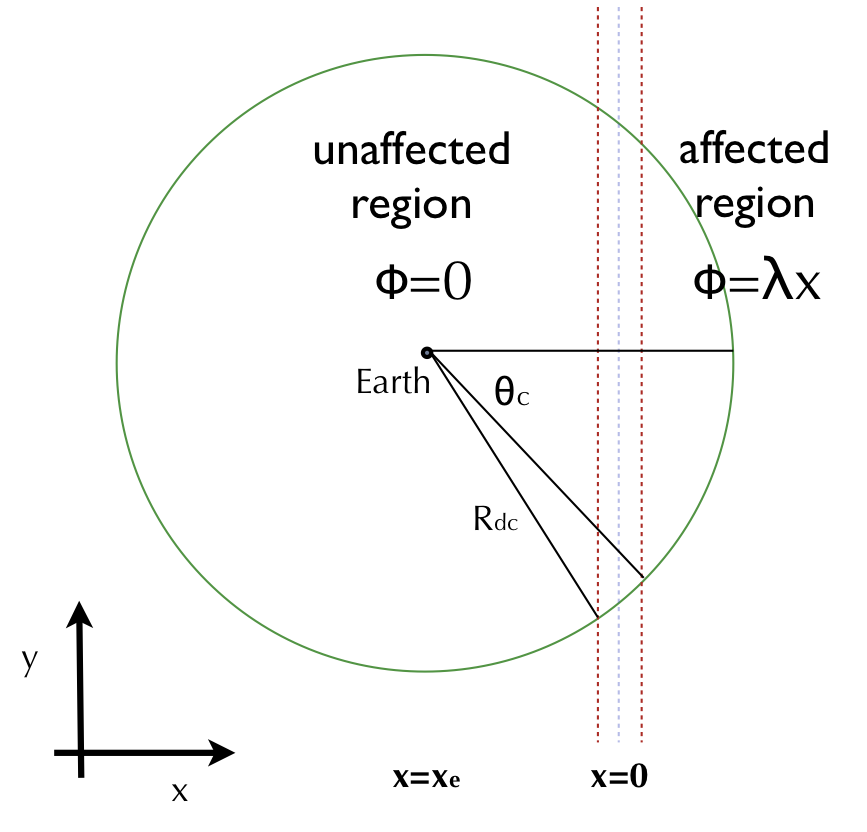}
\caption{\label{lssf}The universe at decoupling, showing the earth's last scattering sphere (our past lightcone at the time of decoupling), the regions affected and unaffected by the collision, and the angular size of the affected disk in the CMB temperature map  $\theta_{c}$.}
\end{figure}

Given a perturbation to the Newtonian potential at reheating $\Phi =  \lambda x \Theta(x)$, one can easily estimate the effect on the CMB temperature today.  The essential points are the following: 

\begin{itemize}

\item  Perturbations that are linear in the spatial coordinates (such as $\Phi=\lambda x$) remain so regardless of the cosmological evolution (to first order in perturbation theory), because the equations that govern the evolution of cosmological perturbations are all second-order in spatial derivatives.

\item The evolution equations are causal---the effects of an initial perturbations are confined to within its light- or sound-cone.

\item By the time of last scattering, the radius of the sound-cone $r_{sh}$ of a perturbation present at reheating corresponds to approximately .6 degrees on the CMB sky ($r_{sh}/D_{dc}\approx.01$).

\end{itemize}

Given this, one can easily determine the temperature perturbation in the CMB.  The ``kink'' at $x=0$ will evolve, but its effects must be confined within a slab of thickness $2 r_{sh}$.  Outside that slab, the perturbation---which was initially either zero or linear in $x$---cannot evolve in shape, although its amplitude can (and does) change.  Therefore the perturbation to the Newtonian potential at $t=t_{dc}$ must be of the following form:
\be \label{phidc}
\Phi(t_{dc}, x,y,z) = \cases{0 ~~~~~~~~~~~~~~~~ ~~~~~~~ x<-r_{sh} \cr f(x)  ~~~~~~~~~~~~ -r_{sh}<x<r_{sh} \cr \lambda x ~~~~~~~ ~~~~ ~~~~~~~~~~~~ x>r_{sh}},
\ee
where $f(x)$ is some function that must be determined by a more careful analysis, but which one may expect to be a smooth deformation of the ``kink" $x \Theta(x)$ (see \figref{lssf}).

To determine the effects of such a perturbation on the CMB in the Sachs-Wolfe approximation, one  needs to project $\Phi(t_{dc}, x, y, z)$ onto the earth's last scattering sphere (of radius $D_{dc}$).  This is trivial: if the earth's comoving location is $x-x_e=y=z=0$ and choosing spherical coordinates centered on the earth with the north pole $\theta=0$  on the $x$-axis, one has $x-x_e = D_{dc} \cos \theta$.  Therefore the effects of the perturbation on the CMB will be confined to the interior of a disk of angular radius 
\be
\theta_c = \cos^{-1}\left( -x_e-r_{sh} \over D_{dc} \right).
\ee 
The effects can be very easily determined everywhere except within the annulus $ \cos^{-1}\left( -x_e+r_{sh} \over D_{dc} \right) < \theta <  \cos^{-1}\left(-x_e- r_{sh} \over D_{dc} \right)$ near the rim of the disk, where it is determined by the function $f(x)$.  The angular width of this annulus is close to $2\times .6^{\circ}=1.2^{\circ}$ for a large radius disk ($|x_{e}| \ll D_{dc}$), and increases for smaller disks.

Inside the inner edge of the annulus, where $\theta <  \cos^{-1}\left( -x_e+r_{sh} \over D_{dc} \right)$, the CMB temperature perturbation in the Sachs-Wolfe approximation is simply linear in $\cos \theta$, as first predicted in \cite{wwc2} (see also \cite{Czech:2010rg}).  The numerical evolution of \cite{kls} confirms that this is an accurate approximation, and that the temperature profile within the annulus is smooth and relatively featureless.

\subsection{CMB polarization}

The free-streaming approximation discussed in the previous subsection is not entirely valid.  Approximately 10\% of CMB photons Thomson re-scatter at $z<z_{dc}$.  This re-scattering leads to a net linear polarization of the CMB radiation if the radiation incident on the scatterer has a non-zero quadrupole moment \cite{Bond:1984fp,Zaldarriaga:1996xe}.  CMB polarization can be decomposed into two types, $E$-mode and $B$-mode.  $B$-mode polarization originates from  tensor perturbations.  Because the bubble perturbation is invariant under rotations around the $x$-axis and does not break chirality, it will not generate $B$-modes.  Another way to understand the same fact is to note that the symmetries of a two bubble collision suffice to prevent the production of gravity waves; see \cite{hms,Kosowsky:1991u}.   

However, the perturbation will generate $E$-mode polarization.  In fact, the pattern of polarization is completely determined by the collision perturbation at decoupling $\Phi(t_{dc},x,y,z)$.  It can be determined accurately using numerical evolution, and as for the temperature this is necessary to determine the detailed structure near the edge of the disk.  In contrast to the temperature perturbation, the polarization signal contains a distinct and interesting substructure within the annulus at the disk's edge, see \cite{Czech:2010rg,kls}  for details.

\paragraph{Analytic approximation for polarization:}  One can estimate the polarization signal using an analytic approximation \cite{Czech:2010rg}.  In this approximation one computes the quadrupole moment of the temperature distribution seen by scattering electrons along the earth's line of sight, making use of the perturbation at decoupling \eqref{phidc} but with $f(x)=\lambda x\Theta(x)$.  The result is that the polarization is non-zero in a wide annulus surrounding the edge of the disk, and has an additional feature confined to a much narrower annulus \cite{Czech:2010rg}.

To understand this result, the key point is to realize that a linear perturbation at decoupling $\Phi(t_dc) = \lambda x$ produces precisely zero quadrupole moment in the radiation incident on a scatterer---it is pure dipole.  Therefore, any scatterer with a last scattering sphere (the scatterer's lightcone at decoupling) that is entirely in the linear region $x>r_sh$ or $x<r_sh$ will not produce linear polarization in the CMB.  Only scatterers with lightcones that intersect $x=0$ at decoupling will see a bath of incident radiation with a non-zero quadrupole component.

Most re-scattering of CMB photons occurs at two times in the history of the universe:  around the time of decoupling $z_{dc} \sim 1000$, and much later at reionization, at $z_{ri} \sim 10$.  The lightcone of a scattering electron at $z \sim 1000$ is much smaller than the lightcone of the earth today---it is roughly $1^{\circ}$ across.  The lighcone of an electron at $z \sim 10$ is by contrast much larger 
Therefore scattering at reionization produces an effect within a broad annulus, while scattering near decoupling produces a sharp feature that must be resolved numerically. 

\subsection{Other cosmological observables}

A number of  cosmological observables other than the CMB will be affected by a bubble collision.  A linear gradient or $x \Theta(x)$ kink in the Newtonian potential $\Phi$ leads to a coherent flow for large-scale structures, a problem that was studied in a radiation dominated toy universe in \cite{ktflow}.  The kink corresponds to a wall of over- or under-density that could potentially be detected in large-scale structure or lensing surveys.  Observations of 21cm radiation could serve as a sensitive probe of large-scale structures with unusual symmetries such as this \cite{Kleban:2007jd}.   If the collision affects light fields other than the inflaton, it could lead to large-scale variations in other observables (for example, if the fine structure constant can vary as in \cite{Moss:2010qa}).   Generally speaking, any measure of the large-scale structure of density perturbations would be sensitive to the effects of a bubble collision.

\section{Probabilities and measures}

The late-time statistics of the clusters of bubbles that form from collisions turns out to be a very rich and interesting topic   \cite{gw,  ggv,Freivogel:2006xu, bubmeas, Bousso:2008as, Garriga:2008ks, fk, Sekino:2009kv, Sekino:2010vc}, but one that goes beyond the scope of this review.  Here, I will discuss only the question of the probability for observing these signals given what we know about cosmology and the underlying microphysics, and what the most probable ranges are for the parameters describing their effects.

To begin to address this, consider an observer inside a bubble expanding into an inflating false vacuum.  To the extent that we can regard the spacetime outside the bubble as unperturbed by its presence, the spacetime outside is pure de Sitter space.  de Sitter spacetime has an event horizon---events separated by more than one Hubble length cannot influence each other.  Therefore the only bubbles that can collide with the observer's bubble are those that nucleate within one false-vacuum Hubble length $1/H_f$ of the wall of the bubble.  

At any given time one can imagine a spherical shell of false vacuum, with area equal to the surface area of the observer's bubble at that time, and thickness one false vacuum Hubble length $1/H_f$.  If a new bubble forms within that shell, it will collide with the observer's.  Therefore we should expect the rate of bubble collisions to be 
\be \label{Ndot}
\langle dN/dT \rangle \sim A(T) H_f^{-1} \Gamma,
\ee
where $T$ is a time coordinate in the false vacuum and  $T=A(T)=0$ is the time of nucleation of the observer's bubble.

To proceed, we need some information about $A(T)$.  The first observation is that since the bubble is expanding, it is a monotonically increasing function of $T$.  Therefore the total number of bubble collisions after infinite time will be infinite.  This is to be expected---the bubble expands eternally, and the rate for collisions does not go to zero (it increases).  As a result, bubbles form infinite clusters \cite{gw}.

\subsection{Observable collisions} 

We are primarily interested in collisions that could be visible to us today.  To see what this implies, recall that bubbles when they form are dominated by negative spatial curvature.  This persists for a physical time interval $t \sim 1/H_i$, where $H_i$ is the Hubble scale during slow roll inflation inside the bubble, because the energy density in curvature redshifts like $a(t)^{-2} \sim t^{-2}$ during curvature domination.  Inflation begins when the energy density in the inflaton potential, $\rho_V \sim H_i^2$, is of order the energy density in curvature $t^{-2}$.

If a given bubble collision is visible to us today, we must be inside the lightcone of the nucleation event of the colliding bubble.  Therefore we should trace back our own past lightcone and determine the area $A$ of our bubble's wall at the moment our past lightcone intersects it.  Collisions which occur later, when $A$ is larger, are not yet visible.  

The details were first worked out correctly in \cite{bubmeas}, but the result is intuitive.  Null rays propagate a comoving distance equal to the conformal time.  Conformal time $\tau$ inside the bubble universe is related to physical time $t$ by $\tau = \int dt/a(t)$.  Inflation makes $a(t)$ exponentially large, which means that during and after inflation very little conformal time passes.  In fact, if there is enough inflation to solve the curvature problem, most of the conformal time inside the bubble is before the beginning of inflation, when the universe is curvature dominated.

During curvature domination, $a(t) \sim t$ and therefore $\tau \sim \ln t$.  Hence, we expect $\tau \sim \ln 1/H_i$.  To complete the estimate, recall that negative spatial curvature means that the areas of spheres grow exponentially with comoving distance:  a sphere of comoving radius $\rho$ has an area of order $\cosh^2 \rho$.  Therefore we can expect the area of our bubble wall at the moment out past lightcone intersects it to be of order $A \sim \cosh^2 \ln(1/H_i) \sim 1/H_i^2$.
Putting this together gives
\be 
\langle dN/dT \rangle \sim H_i^{-2} H_f^{-1} \Gamma = \gamma H_f^3/H_i^2.
\ee

We are not quite finished.  To complete the estimate, we should integrate $dN/dT$ back to $T=0$, the nucleation time of the observer's bubble.  This integral will introduce an additional factor of $H_f$ on the right-hand side.  Perhaps the simplest way to see this is to apply the argument about null ray propagation to the false vacuum---one immediately sees that the lightcones of bubbles travel a comoving distance $1/H_f$ in a time $1/H_f$, and then slow down and stop (in comoving distance) exponentially rapidly.

Finally, we might be interested in bubble collisions with lightcones that divide the part of the last scattering surface we can see today into two pieces, rather than completely encompass it.  The details can be found in \cite{bubmeas}, but taking this into account introduces one more factor, $\sqrt{\Omega_k(t_0)} \sim D_{dc}/a(t_{0})$---the ratio of the comoving size of the earth's decoupling sphere to the radius of spatial curvature today.  

In the end, the result is \cite{bubmeas}
\be
\langle N \rangle \sim \sqrt{\Omega_k} H_i^{-2} H_f^{-2} \Gamma = \gamma \sqrt{\Omega_k} \( H_f/H_i \) ^2.
\ee
The significance of this result is that even though $\gamma$ is small, the ratio $(H_f/H_i)^2$ is expected to be very large.  Indeed, lack of $B$-mode polarization in the CMB indicates that $(M_{Pl}/H_i)^2  \simgeq 10^{10}$.  If $H_f$ is a fundamental scale, one therefore expects this ratio to be very large.  The decay rate $\gamma \sim e^{-S}$ must be small in order for this semi-classical analysis to be valid.  However, in the string theory landscape there are an enormous number of decay channels, and the relevant decay rate is the fastest one available for our parent false vacuum.  Finally, observational constraints are consistent with $\sqrt{\Omega_k} < .1$.

Given the current state of understanding of the landscape and the scale of inflation, it is impossible to make a definite statement---but it doesn't appear that any true fine-tuning is necessary to achieve $\langle N \rangle > 1$.

\subsection{Spot sizes}

Given the symmetries of the parent false vacuum, one would expect the distribution of spot centers to be uniform on the sky ({\it c.f.} \cite{ggv}, but also \cite{bubmeas}).  What about the distribution of sizes?

This can be estimated very easily.  A glance at \eqref{Ndot} shows that the rate of bubble collisions increases with time (since $\dot A > 0$).  Earlier collisions make larger spots on the sky---in fact, very early collisions completely encompass the part of last scattering surface visible to us.  Therefore, we might expect more smaller spots. 

However, this effect is quite small so long as $\Omega_k \ll 1$ today.  The reason is that, at least if we focus only on collisions with lightcones that divide our last scattering sphere, we are looking at collisions that occurred over a time interval where $A(T)$ didn't vary significantly.  As a result, the distribution of collision times over that interval is close to uniform.

The implications for spot sizes are easy to see.  Each collision produces a light ``cone", which at any given time is really a hyperbolic sheet.  In the approximation in which we ignore the curvature of that sheet, it is simply a plane situated at some transverse coordinate $x(t)$.  Uniform distribution of time $T$ corresponds to a uniform distribution of $x$ over the small range ($|x| \simleq \sqrt{\Omega_k}$) that is visible today.  Therefore, we should expect the edges of the regions affected by bubble collisions to be well approximated by randomly situated, randomly oriented planes.  

Since  $x = D_{dc} \cos \theta$, a uniform distribution $dx$ implies a distribution  $d(\cos \theta_c) = \sin \theta d\theta_c$ for the angular radius of the affected spots in the CMB map.  Spots with small angular size are quite rare, as are those that nearly cover the entire sky.  

One interesting implication of this result is that super-horizon ``spots"---collisions with lightcones that encompass our entire last scattering surface---are significantly more common than those that divide it and produce a visible spot.  Such large collisions are difficult to detect, because they are expected to produce a pattern that is primarily dipole and therefore is masked by the earth's peculiar motion.  However, it is important to note that not all effects of such superhorizon perturbations are masked, and some are in principle observable.

\subsection{Spot brightness}

Another important question for the observability of CMB spots is their expected magnitude, parametrized by the size of the temperature perturbation at the center of the spot.  Given a model that produces bubbles and its relation to the inflaton, this question can be answered by following the techniques outlined in this paper  (starting with \eqref{infexp} and a model for the inflationary potential $V(\phi)$) and detailed in the references.  

Even without a model, one should be able to arrive at a reasonable estimate or parametrize the uncertainty.  This was addressed to some extent in \cite{wwc2} and \cite{bubmeas}, which calculated the distribution on de Sitter invariant distances between bubbles (which is one factor that affects the brightness) but the program has not been carried out in detail in any specific model.  

\section{Conclusions}

Very recently, a search of the WMAP temperature data \cite{Feeney:2010dd, Feeney:2010jj} uncovered several cold and hot spots that are consistent with the signal predicted in \cite{wwc2}.  While further analysis is necessary to determine whether or not these features are truly anomalous, this result raises the stakes significantly.  A discovery of a bubble collision in the CMB would confirm a prediction of string theory, and more broadly demonstrate the reality of eternal inflation, revolutionize our understanding of the big bang, and indicate the existence of other ``universes".  With such possibilities in sight, it is well worth pursuing this program vigorously.

\section*{Acknowledgements}
I thank Guido D'Amico, Ben Freivogel, Roberto Gobbetti, Lam Hui, Thomas Levi,  Kris Sigurdson, Alex Vilenkin, and Matias Zaldarriaga for discussions.
My work is supported by NSF CAREER grant PHY-0645435.


\bibliographystyle{utphys}

\bibliography{review}

\end{document}